\documentstyle[amsfonts,preprint,aps]{revtex}
\begin{document}
\title{Realization of the $1\rightarrow 3$ optimal phase-covariant quantum cloning
machine}
\author{Fabio Sciarrino and Francesco De Martini}
\address{Dipartimento di Fisica and \\
Istituto Nazionale per la Fisica della Materia\\
Universit\`{a} di Roma ''La Sapienza'', Roma, 00185 - Italy}
\maketitle

\begin{abstract}
The $1\rightarrow 3$ quantum phase covariant cloning, which optimally clones
qubits belonging to the equatorial plane of the Bloch sphere, achieves the
fidelity ${\cal F}_{cov}^{1\rightarrow 3}=0.833$, l$\arg $er than for the $%
1\rightarrow 3$ universal cloning ${\cal F}_{univ}^{1\rightarrow 3}=0.778$.
We show how the $1\rightarrow 3$ phase covariant cloning can be implemented
by a smart modification of the standard {\it universal} quantum machine by a
projection of the output states over the symmetric subspace. A complete
experimental realization of the protocol for polarization encoded qubits
based on non-linear and linear methods will be discussed.
\end{abstract}


In the last years a great deal of efforts has been devoted to the
realization of the optimal approximations to the quantum cloning and
flipping operations over an unknown qubit $\left| \phi \right\rangle $. Even
if these two processes are unrealizable in their exact forms \cite{Woot82}, 
\cite{Bech99}, they can be optimally approximated by the corresponding
universal machines, i.e., by the universal quantum cloning machine (UQCM)
and the universal-NOT (U-NOT) gate \cite{Buze96}. The optimal quantum
cloning machine has been experimentally realized following several
approaches, i.e. by exploiting the process of stimulated emission in a
quantum-injected optical parametric amplifier (QI-OPA) \cite
{DeMa98,DeMa02,Lama02,Fase02}, by a quantum network \cite{Cumm02} and by
acting with projective operators over the symmetric subspaces of many qubits 
\cite{Ricc04,Irvi04}. The $N\rightarrow M$ UQCM transforms $N$ input qubits
in the state $\left| \phi \right\rangle $ into $M$ entangled output qubits
in the mixed state $\rho _{out}.$ The quality of the resulting copies is
quantified by the fidelity parameter ${\cal F}_{univ}^{N\rightarrow
M}=\left\langle \phi \right| \rho _{out}\left| \phi \right\rangle =\frac{%
N+1+\beta }{N+2}$ with $\beta =\frac{N}{M}\leq 1.$

Not only the perfect cloning of unknown qubit is forbidden but also perfect
cloning of subsets containing non orthogonal states. This no-go theorem
ensures the security of cryptographic protocols as $BB84$ \cite{Gisi02}.
Recently {\it state dependent} cloning machines have been investigated that
are optimal respect to a given ensemble \cite{Brus00}. The partial a-priori
\ knowledge of the state allows to reach a higher fidelity than for the
universal cloning. In particular the $N\rightarrow M$ {\it phase-covariant
quantum cloning machine} (PQCM) considers the cloning of $N$ into $M$ output
qubits, where the input ones belong to the equatorial plane of the
corresponding Poincare' sphere, i.e. expressed by: $\left| \phi
\right\rangle =2^{-1/2}\left( \left| 0\right\rangle +e^{i\phi }\left|
1\right\rangle \right) $. The values of the optimal fidelities ${\cal F}%
_{cov}^{N\rightarrow M}$ for this machine have been found \cite{DAri03}. In
the present article we will restrict ourselves to the case in which $N=1.$
For $M$ assuming odd values it is found ${\cal F}_{cov}^{1\rightarrow M}=%
%TCIMACRO{\UNICODE[m]{0xbc}}%
%BeginExpansion
{\frac14}%
%EndExpansion
\left( 3+M^{-1}\right) \;$while in the case of even $M-$values ${\cal F}%
_{cov}^{1\rightarrow M}=%
%TCIMACRO{\UNICODE[m]{0xbd}}%
%BeginExpansion
{\frac12}%
%EndExpansion
\left( 1+%
%TCIMACRO{\UNICODE[m]{0xbd}}%
%BeginExpansion
{\frac12}%
%EndExpansion
\sqrt{1+2M^{-1}}\right) $. In particular we have ${\cal F}%
_{cov}^{1\rightarrow 2}=0.854$ to be compared with ${\cal F}%
_{univ}^{1\rightarrow 2}=0.833$ and ${\cal F}_{cov}^{1\rightarrow 3}=0.833$
with: ${\cal F}_{univ}^{1\rightarrow 3}=0.778$.

It is worthwhile to enlighten the connections existing between the cloning
processes and the theory of quantum measurement \cite{Brus98}. The concept
of universal quantum cloning is indeed related to the problem of optimal
quantum state estimation \cite{Mass95} since for $M\rightarrow \infty ,$ $%
{\cal F}_{univ}^{N\rightarrow M}\rightarrow {\cal F}_{estim}^{N}=\frac{N+1}{%
N+2}$ where ${\cal F}_{estim}^{N}$ is the optimal fidelity for the state
estimation of any ensemble of $N$ unknown, identically prepared qubits.
Likewise, the phase-covariant cloning has a connection with the estimation
of an equatorial qubit, that is, with the problem of finding the optimal
strategy to estimate the value of the phase $\phi $ \cite{Hole82}, \cite
{Derk98}$.$ Precisely, the optimal strategy consists of a POVM corresponding
to a Von Neumann measurement of $N$ input qubits characterized by a set of $%
N+1$ orthogonal projectors and achieves the fidelity ${\cal F}_{phase}^{N}\ $%
\cite{Derk98}. In general for $M\rightarrow \infty ,$ ${\cal F}%
_{cov}^{N\rightarrow M}\rightarrow {\cal F}_{phase}^{N}$. For $N=1$ is
found: ${\cal F}_{cov}^{1\rightarrow M}={\cal F}_{phase}^{1}+\frac{1}{4M}$
with ${\cal F}_{phase}^{1}=3/4.$

To our knowledge, no PQCM device\ has been implemented experimentally in the
domain of Quantum Optics \cite{Du03,Fiur03}. In the present work we report
the implementation of a $1\rightarrow 3$ PQCM by adopting a modified
standard $1\rightarrow 2$ UQCM and by further projecting the output qubits
over the symmetric subspace \cite{DeMa02,Ricc04}. Let the state of the input
qubit be expressed by: $\left| \phi \right\rangle _{S}=\alpha \left|
0\right\rangle _{S}+\beta \left| 1\right\rangle _{S}$ with {\it real}
parameters $\alpha $ and $\beta $ and $\alpha ^{2}+\beta ^{2}=1$. The output
state of the $1\rightarrow 2$ UQCM device reads: 
\begin{equation}
\left| \Sigma \right\rangle _{SAB}=\sqrt{\frac{2}{3}}\left| \phi
\right\rangle _{S}\left| \phi \right\rangle _{A}\left| \phi ^{\perp
}\right\rangle _{B}-\frac{1}{\sqrt{6}}\left( \left| \phi \right\rangle
_{S}\left| \phi ^{\perp }\right\rangle _{A}+\left| \phi ^{\perp
}\right\rangle _{S}\left| \phi \right\rangle _{A}\right) \left| \phi
\right\rangle _{B}
\end{equation}
The qubits $S$ and $A$ are the optimal cloned qubits while the qubit $B$ is
the optimally flipped one. We perform the operation $U_{B}=\sigma _{Y}$ on
the qubit $B.$ This local flipping transformation of $\left| \phi
\right\rangle _{B}$\ leads to: $\left| \Upsilon \right\rangle _{SAB}=({\Bbb I%
}_{S}\otimes {\Bbb I}_{A}\otimes U_{B})\left| \Sigma \right\rangle _{SAB}=%
\sqrt{\frac{2}{3}}\left| \phi \right\rangle _{S}\left| \phi \right\rangle
_{A}\left| \phi \right\rangle _{B}-\frac{1}{\sqrt{6}}\left( \left| \phi
\right\rangle _{S}\left| \phi ^{\perp }\right\rangle _{A}+\left| \phi
^{\perp }\right\rangle _{S}\left| \phi \right\rangle _{A}\right) \left| \phi
^{\perp }\right\rangle _{B}$. By this non-universal cloning process three 
{\it asymmetric} copies have been obtained: two clones (qubits $S$ and $A)$
with fidelity $5/6$, and a third one (qubit $B$) with fidelity $2/3$. We may
now project $S,$ $A$ and $B$ over the symmetric subspace and obtain three
symmetric clones with a higher average fidelity. The symmetrization operator 
$\Pi _{sym}^{SAB}$ reads as $\Pi _{sym}^{SAB}=\left| \Pi _{1}\right\rangle
\left\langle \Pi _{1}\right| +\left| \Pi _{2}\right\rangle \left\langle \Pi
_{2}\right| +\left| \Pi _{3}\right\rangle \left\langle \Pi _{3}\right|
+\left| \Pi _{4}\right\rangle \left\langle \Pi _{4}\right| $ where $\left|
\Pi _{1}\right\rangle =\left| \phi \right\rangle _{S}\left| \phi
\right\rangle _{A}\left| \phi \right\rangle _{B}$, $\left| \Pi
_{2}\right\rangle =\left| \phi ^{\perp }\right\rangle _{S}\left| \phi
^{\perp }\right\rangle _{A}\left| \phi ^{\perp }\right\rangle _{B}$, $\left|
\Pi _{3}\right\rangle =\frac{1}{\sqrt{3}}\left( \left| \phi \right\rangle
_{S}\left| \phi ^{\perp }\right\rangle _{A}\left| \phi ^{\perp
}\right\rangle _{B}+\left| \phi ^{\perp }\right\rangle _{S}\left| \phi
\right\rangle _{A}\left| \phi ^{\perp }\right\rangle _{B}+\left| \phi
^{\perp }\right\rangle _{S}\left| \phi ^{\perp }\right\rangle _{A}\left|
\phi \right\rangle _{B}\right) $ and $\left| \Pi _{4}\right\rangle =\frac{1}{%
\sqrt{3}}\left( \left| \phi \right\rangle _{S}\left| \phi \right\rangle
_{A}\left| \phi ^{\perp }\right\rangle _{B}+\left| \phi ^{\perp
}\right\rangle _{S}\left| \phi \right\rangle _{A}\left| \phi \right\rangle
_{B}+\left| \phi \right\rangle _{S}\left| \phi ^{\perp }\right\rangle
_{B}\left| \phi \right\rangle _{A}\right) .$ The symmetric subspace has
dimension 4 since three qubits are involved. The probability of success of
the projection is equal to $\frac{8}{9}$. The normalized output state $%
\left| \xi \right\rangle _{SAB}=\Pi _{sym}^{SAB}\left| \Upsilon
\right\rangle _{SAB}$ is 
\begin{equation}
\left| \xi \right\rangle _{SAB}=\frac{\sqrt{3}}{2}\left| \phi \right\rangle
_{S}\left| \phi \right\rangle _{A}\left| \phi \right\rangle _{B}-\frac{1}{2%
\sqrt{3}}\left( \left| \phi \right\rangle _{S}\left| \phi ^{\perp
}\right\rangle _{A}\left| \phi ^{\perp }\right\rangle _{B}+\left| \phi
^{\perp }\right\rangle _{S}\left| \phi \right\rangle _{A}\left| \phi ^{\perp
}\right\rangle _{B}+\left| \phi ^{\perp }\right\rangle _{S}\left| \phi
^{\perp }\right\rangle _{A}\left| \phi \right\rangle _{B}\right)
\label{outputPQCM}
\end{equation}
Let us now estimate the output density matrices of the qubits $S,$ $A$ and $%
B $%
\begin{equation}
\rho _{S}=\rho _{A}=\rho _{B}=\frac{5}{6}\left| \phi \right\rangle
\left\langle \phi \right| +\frac{1}{6}\left| \phi ^{\perp }\right\rangle
\left\langle \phi ^{\perp }\right|  \label{outputdensitymatrices}
\end{equation}
This leads to the fidelity ${\cal F}_{cov}^{1\rightarrow 3}=5/6$ equal to
the optimal one \cite{Brus00,DAri03}.

By applying a different unitary operator $U_{B}$ to the qubit $B$ we can
implement the phase-covariant cloning for different equatorial planes.
Interestingly, note that by this symmetrization technique a depolarizing
channel $E_{dep}(\rho )=%
%TCIMACRO{\UNICODE[m]{0xbc}}%
%BeginExpansion
{\frac14}%
%EndExpansion
\left( \rho +\sigma _{X}\rho \sigma _{X}+\sigma _{Y}\rho \sigma _{Y}+\sigma
_{Z}\rho \sigma _{Z}\right) $ on channel $B$ \ transforms immediately the
non-universal phase covariant cloning into the {\it universal} $1\rightarrow
3$ UQCM with the overall fidelity ${\cal F}_{univ}^{1\rightarrow 3}=$ $7/9$.
This represent a relevant new proposal to be implemented within the $%
1\longrightarrow 2$ UQCM QI-OPA device or other $1\longrightarrow 2$
U-cloning schemes \cite{DeMa02,Ricc05}. Let us return to the $1\rightarrow 3$
PQCM. In the present scheme the input qubit, to be injected into a QI-OPA
over the spatial mode $k_{1}$ with wavelength (wl) $\lambda $, is encoded
into the polarization $(\overrightarrow{\pi })$ state $\left| \phi
\right\rangle _{in}=\alpha \left| H\right\rangle +\beta \left|
V\right\rangle $ of a single photon, where $\left| H\right\rangle $ and $%
\left| V\right\rangle $ stand for horizontal and vertical polarization:
Figure 1. The QI-OPA consisted of a nonlinear (NL) BBO ($\beta $%
-barium-borate), cut for Type II phase matching and excited by a sequence of
UV\ mode-locked laser pulses having wl. $\lambda _{p}$. The relevant modes
of the NL 3-wave interaction driven by the UV pulses associated with mode $%
k_{p}$ were the two spatial modes with wave-vector (wv) $k_{i}$, $i=1,2$,
each one supporting the two horizontal and vertical polarizations of the
interacting photons. The QI-OPA was $\lambda $-degenerate, i.e. the
interacting photons had the same wl's $\lambda =2\lambda _{p}=795nm$. The
NL\ crystal orientation was set as to realize the insensitivity of the
amplification quantum efficiency$\;$to any input state $\left| \phi
\right\rangle _{in}$ i.e. the {\it universality} (U)\ of the ''cloning
machine'' and of the U-NOT gate \cite{DeMa02}.$\ $This key property is
assured by the squeezing hamiltonian $\widehat{H}_{int}=i\chi \hbar \left( 
\widehat{a}_{1\phi }^{\dagger }\widehat{a}_{2\phi \perp }^{\dagger }-%
\widehat{a}_{1\phi \perp }^{\dagger }\widehat{a}_{2\phi }^{\dagger }\right)
+h.c.$ where the field operator $\widehat{a}_{ij}^{\dagger }$ refers to the
state of polarization $j$ $(j=\phi ,\phi ^{\perp })$, realized on the two
interacting spatial modes $k_{i}$ $(i=1,2)$.

Let us consider the injected photon in the mode $k_{1}$ to have any linear
polarization $\overrightarrow{\pi }{\bf =}\phi $. We express this $%
\overrightarrow{\pi }$-state as $\widehat{a}_{1\phi }^{\dagger }\left|
0,0\right\rangle _{k_{1}}=\left| 1,0\right\rangle _{k_{1}}$ where $\left|
m,n\right\rangle _{k_{1}}$ represents a product state with $m$ photons of
the mode $k_{1}$ with polarization $\phi $, and $n$ photons with
polarization $\phi ^{\perp }$. Assume the input mode $k_{2}$ to be in the 
{\it vacuum state }$\left| 0,0\right\rangle _{k_{2}}$. The initial $%
\overrightarrow{\pi }$-state of modes $k_{i}$ reads $\left| \phi
\right\rangle _{in}=\left| 1,0\right\rangle _{k_{1}}\left| 0,0\right\rangle
_{k_{2}}$ and evolves according to the unitary operator $\widehat{{\bf U}}%
\equiv \exp \left( -i\frac{\widehat{H}_{int}t}{\hbar }\right) $. The
1st-order contribution of the output state of the {\it QIOPA} is $\sqrt{%
\frac{2}{3}}\left| 2,0\right\rangle _{k1}\left| 0,1\right\rangle _{k2}-\sqrt{%
\frac{1}{3}}\left| 1,1\right\rangle _{k1}\left| 1,0\right\rangle _{k2}.$ The
above linearization procedure is justified here by the small experimental
value of the {\it gain }$g\equiv \chi t\approx 0.1$. In this context, the
state $\left| 2,0\right\rangle _{k_{1}},$ expressing two photons of the $%
\phi $ mode $k_{1}$ in the $\overrightarrow{\pi }$-state $\phi ,$
corresponds to the state $\left| \phi \phi \right\rangle $ expressed by the
general theory and implies the $L=2$ cloning of the input $N=1$ qubit.\
Contextually with the realization of cloning on mode $k_{1}$,\ the vector $%
\left| 0,1\right\rangle _{k_{2}}$ expresses the single photon state on mode $%
k_{2}$ with polarization $\phi ^{\perp }$, i.e. the {\it flipped} version of
the input qubit. In summary, the qubits $S$ and $A$ are realized by two
single photons propagating along mode $k_{1}$ while the qubit $B$
corresponds to the $\overrightarrow{\pi }$-state of the photon on mode $%
k_{2} $.

The $U_{B}=\sigma _{Y}$ flipping operation on the output mode $k_{2}$,
implemented by means of two $\lambda /2$ waveplates, transformed the QI-OPA$%
\;$output state into$:$ $\left| \Upsilon \right\rangle _{SAB}$=$\sqrt{\frac{2%
}{3}}\left| 2,0\right\rangle _{k1}\left| 1,0\right\rangle _{k2}-\sqrt{\frac{1%
}{3}}\left| 1,1\right\rangle _{k1}\left| 0,1\right\rangle _{k2}$. The
physical implementation of the projector $\Pi _{sym}^{SAB}$ on the three
photons $\overrightarrow{\pi }$-states was carried out by linearly
superimposing the modes $k_{1}$ and $k_{2}$ on the $50:50$ beam-splitter $%
BS_{A}$ and then by selecting the case in which the 3 photons \ emerged from 
$BS_{A}$ on the same output mode $k_{3}$ (or, alternatively on $k_{4}$) \cite
{Ricc04}. The $BS_{A}$ input-output mode relations are expressed by the
field operators: $\widehat{a}_{1j}^{\dagger }$= $2^{-1/2}(\widehat{a}%
_{3j}^{\dagger }+i\widehat{a}_{4j}^{\dagger })$; $\widehat{a}_{2j}^{\dagger
} $= $2^{-1/2}(i\widehat{a}_{3j}^{\dagger }+\widehat{a}_{4j}^{\dagger })$
where the operator $\widehat{a}_{ij}^{\dagger }$ refers to the mode $k_{i}$
with polarization $j$. The input state of $BS_{A}$ can be re-written in the
following form $\frac{1}{\sqrt{3}}\left( \widehat{a}_{1\phi }^{\dagger 2}%
\widehat{a}_{2\phi }^{\dagger }-\widehat{a}_{1\phi }^{\dagger }\widehat{a}%
_{1\phi \perp }^{\dagger }\widehat{a}_{2\phi \perp }^{\dagger }\right)
\left| 0,0\right\rangle _{k1}\left| 0,0\right\rangle _{k2}.$ By adopting the
previous relations and by considering the case in which 3 photons emerge
over the mode $k_{3},$ the output state is found to be $\frac{1}{2\sqrt{2}}(%
\widehat{a}_{3\phi }^{\dagger 3}+\widehat{a}_{3\phi }^{\dagger }\widehat{a}%
_{3\phi \perp }^{\dagger 2})\left| 0,0\right\rangle _{k3}=\frac{\sqrt{3}}{2}%
\left| 3,0\right\rangle _{k3}+\frac{1}{2}\left| 1,2\right\rangle _{k3}$. The
output fidelity is ${\cal F}_{cov}^{1\rightarrow 3}=\frac{5}{6}.$

Interestingly, the same overall state evolution can also be obtained, with
no need of \ the final $BS_{A}$ symmetrization, at the output of a QI-OPA
with a type II crystal working in a {\it collinear} configuration, as
proposed by \cite{DeMartL98}. In this case the interaction Hamiltonian $%
\widehat{H}_{coll}=i\chi \hbar \left( \widehat{a}_{H}^{\dagger }\widehat{a}%
_{V}^{\dagger }\right) +h.c.$ acts on a single spatial mode $k$. A
fundamental physical property of $\widehat{H}_{coll}$ consists of its
rotational invariance under $U(1)$ transformations, that is, under any
arbitrary rotation around the $z$-axis . Indeed $\widehat{H}_{coll}$ can be
re-expressed as $\frac{1}{2}i\chi \hbar e^{-i\psi }\left( \widehat{a}_{\psi
}^{\dagger 2}-e^{i2\psi }\widehat{a}_{\psi \perp }^{\dagger 2}\right) +h.c.$
for $\psi \in (0,2\pi )$ where $\widehat{a}_{\psi }^{\dagger }=2^{-1/2}(%
\widehat{a}_{H}^{\dagger }+e^{i\psi }\widehat{a}_{V}^{\dagger })$ and $%
\widehat{a}_{\psi \perp }^{\dagger }=2^{-1/2}(-e^{-i\psi }\widehat{a}%
_{H}^{\dagger }+\widehat{a}_{V}^{\dagger })$. Let us consider an injected
single photon with $\overrightarrow{\pi }$-state $\left| \psi \right\rangle
_{in}=2^{-1/2}(\left| H\right\rangle +e^{i\psi }\left| V\right\rangle
)=\left| 1,0\right\rangle _{k}.$ The first contribution to the amplified
state, $\sqrt{6}\left| 3,0\right\rangle _{k}-\sqrt{2}e^{i2\psi }\left|
1,2\right\rangle _{k}$ is identical to the output state obtained with the
device dealt with in the present work up to a phase factor which does not
affect the fidelity value.

The UV pump beam with wl $\lambda _{p}$, back reflected by the spherical
mirror $M_{p}$ with 100\% reflectivity and $\mu -$adjustable position ${\bf Z%
}$, excited the NL crystal in both directions $-k_{p}$ and $k_{p}$, i.e.
correspondingly oriented towards the right hand side and the left hand side
of Fig.1. A Spontaneous Parametric Down Conversion (SPDC) process excited by
the $-k_{p}$ UV\ mode created {\it singlet-states} of photon polarization $(%
\overrightarrow{\pi })$. The photon of each SPDC pair emitted over the mode $%
-k_{1}$ was back-reflected by a spherical mirror $M$ into the NL crystal and
provided the $N=1$ {\it quantum injection} into the OPA excited by the UV
beam associated with the back-reflected mode $k_{p}$. The twin SPDC\ photon
emitted over mode $-k_{2}$ , selected by the ''state analyzer'' consisting
of the combination (Wave-Plate + Polarizing Beam Splitter: $WP_{T}\ $+ $%
PBS_{T}$) and detected by $D_{T}$, provided the ''trigger'' of the overall
conditional experiment. Because of the EPR non-locality of the emitted
singlet, the $\overrightarrow{\pi }$-selection made on $-k_{2}$ implied
deterministically the selection of the input state $\left| \phi
\right\rangle _{in}$on the injection mode $k_{1}$. By adopting a $\lambda /2$
wave-plate ($WP_{T}$) with different orientations of the optical axis, the
following $\left| \phi \right\rangle _{in}$states were injected:$\;\left|
H\right\rangle $ and $2^{-1/2}(\left| H\right\rangle +\left| V\right\rangle
)=\left| +\right\rangle $. A more detailed description of the {\it QI-OPA}
setup can be found in \cite{DeMa02}. The $U_{B}=\sigma _{Y}$ flipping
operation was implemented by two $\lambda /2$ waveplates (wp), as said. The
device $BS_{A}$ was positioned onto a motorized translational stage: the
position $X=0$ in Fig. 2 was conventionally assumed to correspond to the
best overlap between the interacting photon wavepackets which propagate
along $k_{1}$ and $k_{2}$.

The output state on mode $k_{3}$ was analyzed by the setup shown in the
inset of Fig. 1: the field on mode $k_{4}$ was disregarded, for simplicity.
The polarization state on mode $k_{3}$ was analyzed by the combination of
the $\lambda /2$ wp $WP_{C}$ and of the polarizer beam splitter $PBS_{C}$.
For each input $\overrightarrow{\pi }$-state $\left| \phi \right\rangle _{S}$%
, two different measurements were performed. In a first experiment $WP_{C}$
was set in order to make $PBS_{C}$ to transmit $\left| \phi \right\rangle $
and reflect $\left| \phi ^{\bot }\right\rangle .$ The cloned state $\left|
\phi \phi \phi \right\rangle $ was detected by a coincidence between the
detectors $[D_{C}^{1},D_{C}^{2},D_{C}^{3}]$ while the state $\left| \phi
\phi \phi ^{\bot }\right\rangle $, in the ideal case not present, was
detected by a coincidence recorded either by the $D$ set $%
[D_{C}^{1},D_{C}^{2},D_{C}^{\ast }]$, or by $[D_{C}^{1},D_{C}^{3},D_{C}^{%
\ast }]$, or by $[D_{C}^{2},D_{C}^{3},D_{C}^{\ast }]$. In order to detect
the contribution due to $\left| \phi \phi ^{\bot }\phi ^{\bot }\right\rangle 
$, $WP_{C}$ was rotated in order to make $PBS_{C}$ to transmit $\left| \phi
^{\bot }\right\rangle $ and reflect $\left| \phi \right\rangle $ and by
recording the coincidences by one of the sets $[D_{C}^{1},D_{C}^{2},D_{C}^{%
\ast }]$, $[D_{C}^{1},D_{C}^{3},D_{C}^{\ast }]$, $%
[D_{C}^{2},D_{C}^{3},D_{C}^{\ast }]$. The different overall quantum
efficiencies have been taken into account in the processing of the
experimental data. The precise sequence of the experimental procedures was
suggested by the following considerations. Assume the cloning machine turned
off, by setting the optical delay $\left| Z\right| >>c\tau _{coh}$, i.e., by
spoiling the temporal overlap between the injected photon and the UV\ pump
pulse. In this case since the states $\left| \phi \phi \right\rangle $ and $%
\left| \phi ^{\bot }\phi ^{\bot }\right\rangle $ are emitted with same
probability by the machine, the rate of coincidences due to $\left| \phi
\phi \phi \right\rangle $ and $\left| \phi \phi ^{\bot }\phi ^{\bot
}\right\rangle $ were expected to be equal. By turning on the PQCM, i.e., by
setting $\left| Z\right| <<c\tau _{coh}$, the output state (\ref{outputPQCM}%
) was realized showing a factor $R=3$ enhancement of the counting rate of $%
\left| \phi \phi \phi \right\rangle $ and {\it no} enhancement of $\left|
\phi \phi ^{\bot }\phi ^{\bot }\right\rangle $. \ In Fig.2 the coincidences
data for the different state components are reported versus the delay $Z$
for the two input qubits $\left| \phi \right\rangle _{in}$. We may check
that the phase covariant cloning process affects only the $\left| \phi \phi
\phi \right\rangle $ component, as expected. Let us label by the symbol $h$
the output state components as follows: $\left\{ h=1\leftrightarrow \left|
\phi \phi ^{\bot }\phi ^{\bot }\right\rangle \text{, }2\leftrightarrow
\left| \phi \phi \phi ^{\bot }\right\rangle \text{, }3\leftrightarrow \left|
\phi \phi \phi \right\rangle \right\} $. For each index $h$, $b_{h}$ is the
average coincidence rate when the cloning machine is turned off , i.e. $%
\left| Z\right| >>c\tau _{coh}$. while the signal-to-noise (S/N)\ parameter $%
R_{h}$ is the ratio between the peak values of the coincidence rates
detected respectively for $Z\simeq 0$ and $\left| Z\right| >>c\tau _{coh}$.
The optimal values obtained by the above analysis are: $R_{3}=3,$ $R_{1}=1$, 
$b_{3}=b_{1}$ and $b_{2}=0$, $R_{2}=0$. These last values, $h=2$ are
considered since they are actually measured in the experiment:\ Fig.2. The
fidelity has been evaluated by means of the expression ${\cal F}%
_{cov}^{1\rightarrow 3}(\phi )$ = $(3b_{3}R_{3}+2b_{2}R_{2}+b_{1}R_{1})%
\times (3b_{3}R_{3}+3b_{2}R_{2}+3b_{1}R_{1})^{-1}$and by the experimental
values of $b_{h}$, $R_{h}$. For $\left| \phi \right\rangle _{in}=\left|
H\right\rangle $ and $\left| \phi \right\rangle _{in}=\left| +\right\rangle $
we have found respectively $R_{3}=2.00\pm 0.12$ and $R_{3}=1.92\pm 0.06$
(see Fig.2). We have obtained ${\cal F}_{cov}^{1\rightarrow 3}(\left|
+\right\rangle )=0.76\pm 0.01$, and ${\cal F}_{cov}^{1\rightarrow 3}(\left|
H\right\rangle )=0.80\pm 0.01,$ to be compared with the theoretical value $%
0.83$. The fidelity of the cloning $\left| H\right\rangle $ is slightly
increased by a contribution $0.02$ due to an unbalancement of the
Hamiltonian terms.

For the sake of completeness, we have carried out an experiment setting the
pump mirror in the position $Z\simeq 0$ and changing the position $X$ ob $%
BS_{A}$. The injected state was $\left| \phi \right\rangle _{in}=\left|
+\right\rangle $. Due to quantum interference, the coincidence rate was
enhanced by a factor $V^{\ast }$ moving from the position $\left| X\right|
>>c\tau _{coh}$ to the condition $X\approx 0$ . The $\left| \phi \phi
^{\perp }\phi ^{\perp }\right\rangle $ enhancement was found $V_{\exp
}^{\ast }=1.70\pm 0.10$, to be compared with the theoretical value $V^{\ast
}=2$ while the enhancement of the term $\left| \phi \phi \phi \right\rangle $
was found $V_{\exp }^{\ast }=2.16\pm 0.12$, to be compared with the
theoretical value $V^{\ast }=3$. These results, not reported in Fig. 2, are
a further demonstration of the 3-photon interference in the Hong-Ou-Mandel
device.

In conclusion, we have implemented the optimal quantum triplicators for
equatorial qubits. The present approach can be extended in a straightforward
way to the case of $1\rightarrow M$ PQCM for $M$ odd. The results are
relevant in the modern science of quantum communication as the PQCM is
deeply connected to the optimal eavesdropping attack at {\it BB84} protocol,
which exploits the transmission of quantum states belonging to the $x-z$
plane of the Bloch sphere. \cite{Fuch97,Gisi02}. The optimal fidelities
achievable for equatorial qubits are equal to the ones considered for the
four states adopted in $BB84$ \cite{DAri03}. In addition, the phase
covariant cloning can be useful to optimally perform different quantum
computation tasks adopting qubits belonging to the equatorial subspace \cite
{Galv00}.

This work has been supported by the FET European Network on Quantum
Information and Communication (Contract IST-2000-29681: ATESIT), by Istituto
Nazionale per la Fisica della Materia (PRA\ ''CLON'')\ and by Ministero
dell'Istruzione, dell'Universit\`{a} e della Ricerca (COFIN 2002).

\centerline{\bf Figure Captions}

\vskip 8mm

\parindent=0pt

\parskip=3mm

Figure.1. Schematic diagram of phase-covariant cloner, PQCM\ made up by a
QI-OPA and a Hong-Ou-Mandel interferometer $BS_{A}$. INSET: measurement
setup used for testing the cloning process.

Figure.2. Experimental results of the PQCM\ for the input qubits $\left|
H\right\rangle $ and $\left| +\right\rangle =2^{-1/2}(\left| H\right\rangle
+\left| V\right\rangle )$. The measurement time of each 4-coincidence
experimental datum was $\ \sim 13000$ $s.$ The different overall detection
efficiencies have been taken into account. The solid line represents the
best Gaussian fit.

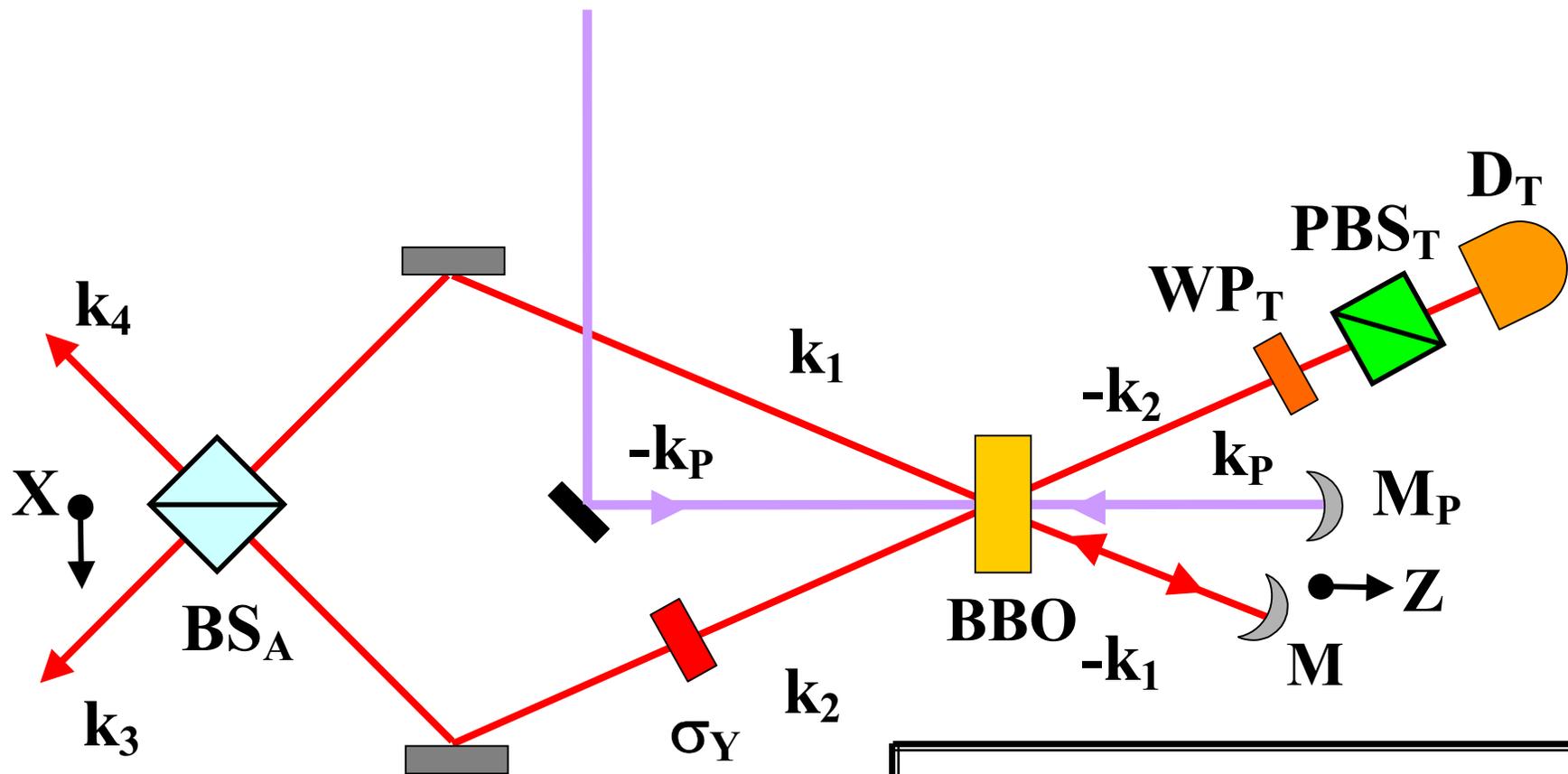
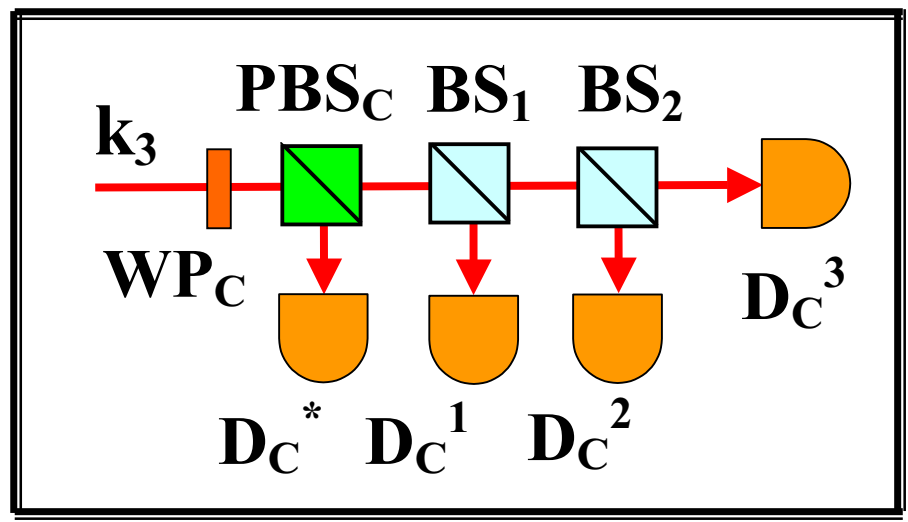
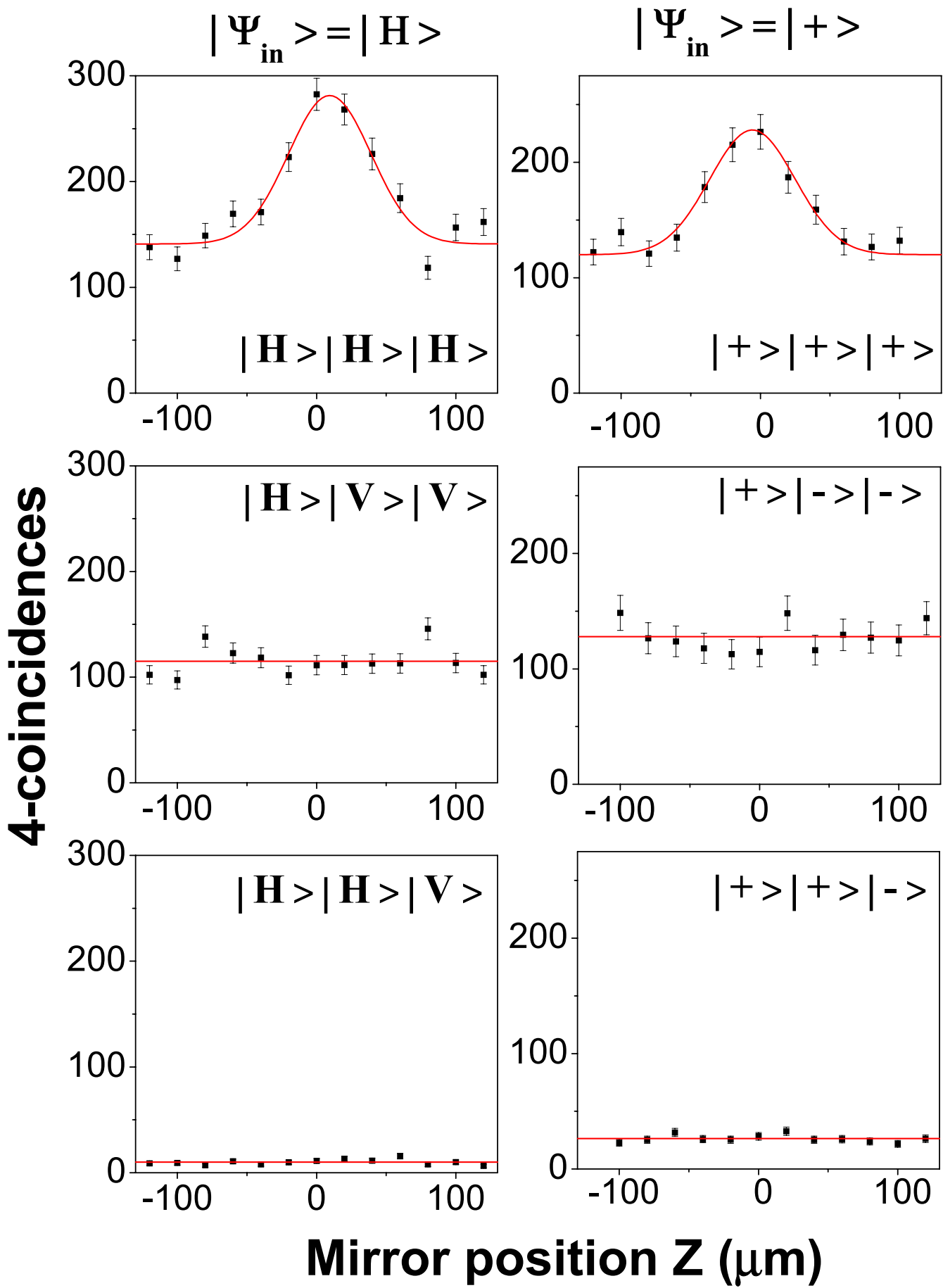
\end{document}